# Giant Nonlocality near the Dirac Point in Graphene

D. A. Abanin[1,2], S. V. Morozov[1,6], L. A. Ponomarenko[1], R. V. Gorbachev[1], A. S. Mayorov[1], M. I. Katsnelson[3], K. Watanabe[4], T. Taniguchi[4], K. S. Novoselov[1], L. S. Levitov[5*], A. K. Geim[1*]

[1]Manchester Centre for Mesoscience & Nanotechnology, University of Manchester, Manchester M13 9PL, UK
[2]Princeton Center for Theoretical Science, Princeton University, Princeton New Jersey 08544, USA
[3]Theory of Condensed Matter, Radboud University Nijmegen, 6525 AJ Nijmegen, Netherlands
[4]National Institute for Materials Science, 1-1 Namiki, Tsukuba, 305-0044 Japan
[5]Department of Physics, Massachusetts Institute of Technology, Cambridge Massachusetts 02139, USA
[6]Institute for Microelectronics Technology, Chernogolovka 142432, Russia

*Transport measurements have been a powerful tool for uncovering new electronic phenomena in graphene. We report nonlocal measurements performed in the Hall bar geometry with voltage probes far away from the classical path of charge flow. We observe a large nonlocal response near the Dirac point in fields as low as 0.1T, which persists up to room temperature. The nonlocality is consistent with the long-range flavor currents induced by lifting of spin/valley degeneracy. The effect is expected to contribute strongly to all magnetotransport phenomena near the neutrality point.*

Graphene continues to attract intense interest, especially as a new electronic system in which charge carriers are Dirac-like particles with linear dispersion and zero rest mass. Transport measurements in graphene have unveiled a number of unique phenomena including two new types of the quantum Hall effect (QHE), minimum metallic conductivity, bipolar superconductivity and Klein scattering (*1-4*). In a number of experiments unusual behavior was found at low temperatures $T$ and high magnetic fields $B$ near the so-called Dirac or neutrality point (NP) where charge career density $n$ tends to zero (*5-9*). However, the NP is also hardest to access experimentally because of charge inhomogeneity (electron-hole puddles) and limited carrier mobilities $\mu$. Furthermore, the fundamental neutral degrees of freedom in graphene, such as spin and valley, evade detection by the standard electrical measurement techniques, even in the best quality samples (here the valley degree of freedom refers to the inequivalence of the pair of conical valence/conduction bands in the Brillouin zone which touch at Dirac points).

In this work we perform nonlocal measurements, previously used to probe the dynamics of population imbalance for edge modes in quantum Hall systems (*10,11*) as well as spin diffusion (*12*) and magnetization dynamics (*13*). The advantage of nonlocal measurements is that they allow one to filter out the ohmic contribution resulting from charge flow and, in doing so, detect more subtle effects that otherwise can remain unnoticed (*10-14*). The measurements were carried out by using more than 20 devices of two different types. Type I devices were made in the conventional way, with graphene placed on top of an oxidized Si wafer (*1-7*), hereafter referred to as GSiO. Type II devices contained thin crystals of hexagonal boron nitride placed between graphene and $SiO_2$ (*15*) (referred to as GBN). All the devices were made in the Hall bar geometry by following the microfabrication procedures described previously (*1,6,15-17*). The GSiO devices had mobility $\mu$ ~10,000 $cm^2$/Vs whereas GBN devices showed much higher $\mu$, between 50,000 and 150,000 $cm^2$/Vs for carrier concentrations $n$ ~$10^{11}$ $cm^{-2}$ (*17*). Typical charge inhomogeneity $n_0$ estimated from the rounding of the conductivity minimum was ~$10^{10}$ and $10^{11}$ $cm^{-2}$ for GBN and GSiO devices, respectively. All our samples exhibited a qualitatively similar nonlocal response; however, its absolute value was 10 to 100 times larger in GBN samples. Unless stated explicitly, the results described below refer equally to both device types.

Figure 1A shows a representative GSiO device, used below to describe different measurement geometries. In the standard Hall bar geometry such that current $I_{14}$ flows between contacts 1 and 4 and voltage $V_{23}$ is measured between contacts 2 and 3, the longitudinal resistivity $\rho_{xx}$ (calculated as $(w/L) \cdot R_{23,14}$ where $L$ and $w$ are the length and width of the Hall bar, and $R_{23,14} = V_{23}/I_{14}$) shows the standard QHE behavior for monolayer graphene with wide regions of zero $\rho_{xx}$ accompanied by well-defined plateaus in Hall resistivity $\rho_{xy}$ (Fig. 1B; fig. S1).

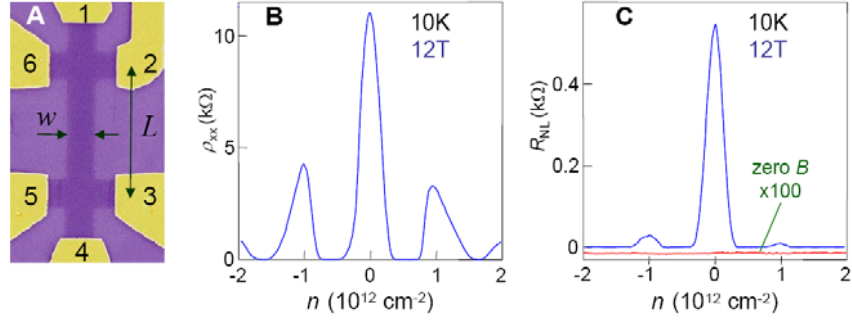

Fig. 1. Local and nonlocal geometries. (**A**) – Electron micrograph (false color) of a GSiO device. The width $w$ =1μm and length $L$ of the Hall bar are indicated. (**B**) – Longitudinal resistivity $\rho_{xx}$ as a function of carrier density $n$ in a perpendicular $B$ =12 T. (**C**) – In the nonlocal geometry, no signal can be detected in zero $B$ (red curve is downshifted for clarity and magnified). Magnetic field gives rise to large nonlocal response $R_{NL}$ shown for standard-quality devices (GSiO type). To assure no contribution from inductive coupling and thermopower, we have used both dc and low-frequency ac measurements with typical driving currents $I$ of 1μA. $R_{NL}$ was confirmed to be independent of $I$ by varying it over 2 orders of magnitude.

In the following, we focus on the nonlocal resistance, $R_{NL}$. The measured signal (e.g., $R_{35,26}$ in Fig. 1C) cannot be understood in terms of the classical picture of charge flow. Indeed, a fraction of applied current $I_{26}$ which flows sideways and reaches the remote region between contacts 3 and 5, is exponentially small in the separation $L$. Using the van der Pauw formalism (*18*), it is straightforward to show that the expected Ohm's law contribution to $R_{NL}$ behaves as $\approx \rho_{xx} \exp(-\pi L/w)$ for both zero and non-zero $B$ (*17*). For our devices, $L$ ranged from 3 to 15 μm and $w$ between 1 and 2 μm. For a typical $L/w$ =5, this translates into minute $R_{NL}$ <$10^{-3}$Ω. In agreement with this estimate, $R_{NL}(B=0)$ was indistinguishable from zero at our maximum resolution (Fig. 1C).

The situation changes radically in finite $B$: $R_{NL}$ remains zero at zeros of $\rho_{xx}$ but between the QHE zeros it can reach values of ~1kΩ, even in the conventional GSiO devices, and exhibits the same overall oscillating pattern as $\rho_{xx}$ (Fig.1C). Although the pattern always remained the same, the amplitude of the nonlocal response varied significantly for different devices. In particular, $R_{NL}$ depended on an exact contact configuration (that is, $R_{35,26} \neq R_{34,26}$), yet with the Onsager relation $R_{35,26}(B) \neq R_{26,35}(B) = R_{35,26}(-B)$ satisfied (fig. S3). $R_{NL}$ was found to become smaller with increasing $L$ and in the presence of extra leads between current and voltage contacts (fig. S3). The strong sample and contact dependence did not allow us to quantify the spatial scale involved in the nonlocality but it can be estimated as exceeding $L$ (that is, ~10μm) in $B$ >5T and $T$ <100K. To emphasize the importance of nonlocal transport near the NP, in (*17*) we describe the standard Hall measurements in two configurations $R_{35,42}$ and $R_{35,46}$, where the same voltage probes were used and the only difference was the swap of one of the current leads. In a classical conductor, this should cause no effect whatsoever but, in graphene, nonlocal transport leads to profound differences between the two supposedly equivalent measurements (fig. S1).

To elucidate the origin of the unexpected nonlocality at the NP, we studied its $T$ and $B$ dependence. The peaks in $R_{NL}$ at filling factors ν =4 and 8 completely disappear above 70K, simultaneously with the disappearance of the zeros in $\rho_{xx}$. Therefore, the nonlocality at ν =4, 8 can be attributed to the standard QHE edge state transport (*10,11*). In contrast, the nonlocal signal at the NP (ν =0) is found to be much more robust (Fig. 2), extending well beyond the QHE regime, into the regime where even Shubnikov-de Haas oscillations are completely absent. At 300K, the nonlocality remains quite profound, with $R_{NL}$ ~1 kΩ at several T and a remnant signal observable in $B$ <<1T. This behavior implies that the nonlocality at the NP occurs via a mechanism, different from the QHE edge state transport (*10,11,17*).

Fig. 2C reveals two temperature regimes. At high $T$, $R_{NL}$ decreases slowly with increasing $T$, whereas below ~30K one can see a rapid increase in $R_{NL}$. The latter correlates with an increase in $\rho_{xx}$ for GBN devices and can be attributed to the onset of an energy gap that opens at ν =0 at low $T$ (*5,7,9,15*). By using the Corbino geometry, we found that the gap did not exceed 20K at 12T for GSiO (*17*). Similar values were reported by other groups (*7,19*). For certain gapped states, the nonlocality can arise due to counter-circulating edge states (*6*). To test this possibility, we carried out nonlocal measurements on devices patterned to have a channel widening that increased devices' edge length more than tenfold, while $L$ between the current and voltage contacts remained the same (*17*). No significant difference in $R_{NL}$ was observed in such devices as compared to those with no widening. This and other observations described in (*17*) provide evidence against edge transport and suggest a bulk transport mechanism even in the low-$T$ gapped state. This conclusion is also consistent with the insulating behavior found in the previous magnetotransport studies at the NP (*5,7,9,15*). The observed sharp increase in $R_{NL}$ at low $T$ (Fig. 2C) may indicate that the dominant nonlocality mechanism changes as the system goes into the gapped state.

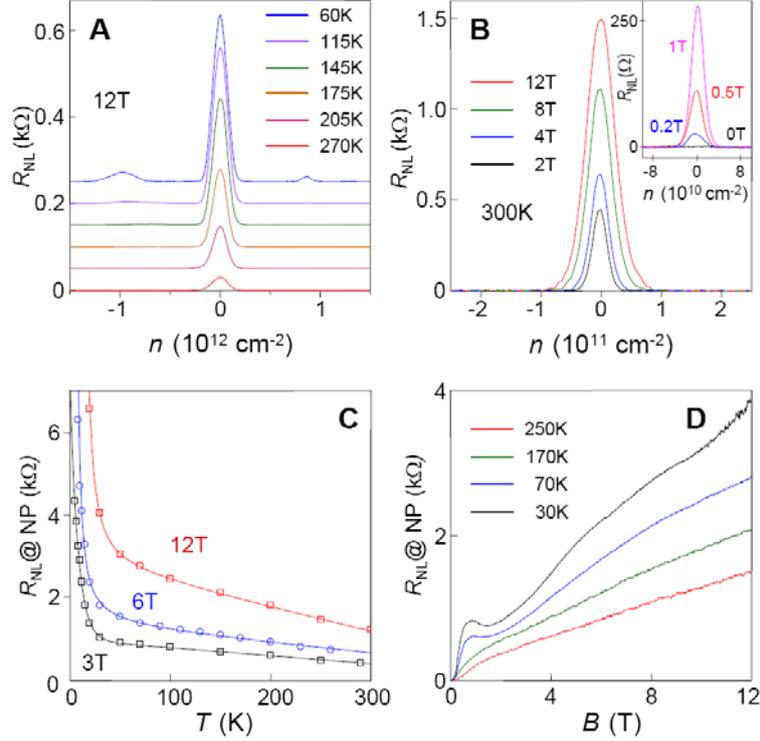

Fig. 2. Nonlocal transport in graphene. (**A**) – $R_{NL}$ for the GSiO device in Fig. 1 at different $T$. In high $B$, the nonlocality at ν =4 persists up to liquid nitrogen $T$. The nonlocal signal at the NP is even more robust with increasing $T$. (**B**) – room-$T$ $R_{NL}$ for a GBN device with μ ≈140,000 cm$^2$/Vs, and with nonlocal voltage contacts separated from the current path by $L$ =5μm. The inset magnifies remnant $R_{NL}$ in small $B$. Even at 0.1T, $R_{NL}$ remains substantial (~10Ω). GSiO devices exhibit a qualitatively similar behavior but with room-$T$ values of $R_{NL}$ ~100 times smaller (*17*). (**C,D**) – $R_{NL}$ at the NP as a function of $T$ for several values of $B$ and as a function of $B$ for several values of $T$, respectively. The data are for the same GBN device as in (**B**). Solid curves in (**C**) are guides to the eye.

Below we discuss the high-$T$ regime, where the gap opening at the NP is irrelevant as no nonlocal signal could be detected even at ν =4 and 8 despite cyclotron gaps being large (~1,000K). The nonlocality observed at high $T$ and low $B$ calls for a quasiclassical explanation that does not involve Landau quantization. At the same time, one has to find a mechanism that naturally extends into the low-$T$ regime where the observed nonlocality becomes increasingly more profound. One possible explanation is the flavor Hall effect (FHE), a bulk mechanism in which nonlocality is mediated by neutral excitations, such as spin and valley flavors, and which works in both quasiclassical and QHE regimes, providing a natural explanation for our experimental findings (*17*).

The basic physics of the FHE is illustrated in Fig. 3 that for simplicity refers to the case of spin. The Zeeman splitting shifts the Dirac cones for opposite spin projections relative to each other. At the NP the spin splitting produces a finite concentration of electrons with spin-up (↑) and holes with spin-down (↓) (Fig. 3A). When electric current is applied, the Lorentz force creates opposite spin-up and spin-down currents, leading to a spatial spin imbalance at zero net Hall voltage at the NP (Fig. 3B). The phenomenology is similar to the spin Hall effect (SHE) resulting from spin-orbit interaction (*20-22*), yet our SHE effect relies on the Zeeman splitting induced by $B$ and occurs in the absence of spin-orbit interaction. In graphene, the SHE can generate long-range spin currents, due to slow spin relaxation (*2,23*), and produce a nonlocal voltage at a remote location via a reverse SHE (*23*), as illustrated in Fig. 3B.

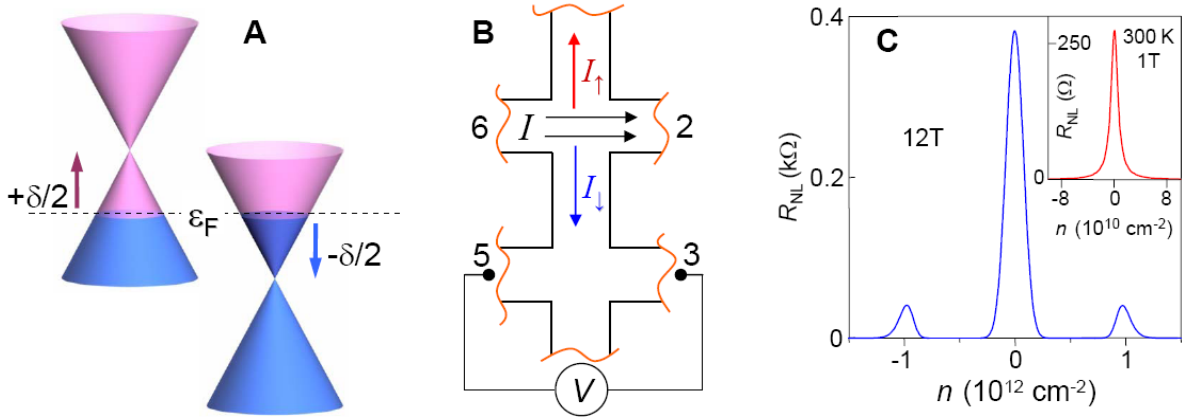

Fig. 3. Spin Hall effect in graphene and nonlocal transport mediated by spin diffusion. (**A**) – Zeeman splitting at charge neutrality produces two pockets filled with electron and holes having opposite spin. (**B**) – In the presence of the Lorentz force, $I$ gives rise to transverse spin currents $I_\uparrow$ and $I_\downarrow$. Because the force has opposite signs for electrons and holes, the net charge current is zero, whereas the net spin current is nonzero. The resulting imbalance in the up/down spin distribution can reach remote regions and generate a voltage drop $V$. (**C**) – Modeled $R_{NL}$ for the QHE regime (main panel) and the quasiclassical regime (inset). The best-fit parameters $n_0 = 4 \times 10^9$ cm$^{-2}$ and Landau level broadening $\Gamma = 200K$ are typical for GBN and GSiO devices, respectively. $R_{NL}$ grows with decreasing $n_0$ and $\Gamma$ (*17*), which is consistent with much larger $R_{NL}$ measured in our GBN devices.

Figure 3C plots the modeled SHE behavior for $R_{NL}$ in GSiO, which captures the main features of the experimental data, most importantly the peak at the NP in $R_{NL}(n)$. The model also predicts maximum value $R_{NL}$ $\sim h/4e^2$, which corresponds to a cut-off due to Landau level broadening (*17*). Such values are indeed in agreement with our measurements in GBN devices (Fig. 2C). The $T$ and $B$ dependences predicted from the simple model are in qualitative agreement with the experiment. The agreement can be further improved by taking into account valley splitting that can give rise to neutral valley currents and additional nonlocality (*17*). In particular, the onset of the valley splitting due to interaction effects (*19*) may be responsible for the observed increase in $R_{NL}$ below 30K. Although our measurements do not probe flavor currents directly, the indirect evidence is overwhelming. The nonlocal phenomena are very rare and, given that we have ruled out edge state transport mechanisms, we believe that the spin/valley Hall effect is the only remaining explanation for our findings.

In conclusion, the profound nonlocality is an essential attribute of electron transport in graphene. The nonlocality is consistent with neutral currents generated by the SHE at high $T$ and, possibly, by the valley Hall effect at liquid-helium $T$. Nonlocal transport, being directly sensitive to neutral degrees of freedom, provides valuable information inaccessible by conventional electrical measurements.

*Acknowledgement* - We thank Daniel Elias, Peter Blake, Ernie Hill, Svetlana Anissimova, Fred Schedin and Irina Grigorieva for their help. This work was supported by the Engineering and Physical Research Council (UK), the Royal Society, U.S. Office of Naval Research, U.S. Air Force Office of Scientific Research and the Körber foundation.

SUPPORTING MATERIAL
*Giant Nonlocality and Spin Hall Effect near the Dirac Point in Graphene*

#### #1. Influence of the nonlocality on local measurements

Figure S1 shows two sets of Hall measurements by using the same voltage contacts (3 and 5) and changing only one of the current contacts (contacts 2 and 6 are swapped in the measurements shown in panels **a** and **b**). At first glance, Hall resistivity $R_{xy}$ looks more or less the same but further analysis shows that the traces differ by as much as 500 Ohms. Indeed, panel **c** plots the difference between $R_{xy}$ shown in **a** and **b**. The dip around +12V can be explained by the nonlocal edge-state transport [S1,S2]. The measurements are expected to be electron-hole symmetric but a similar dip on the hole side is smeared by charge inhomogeneity.

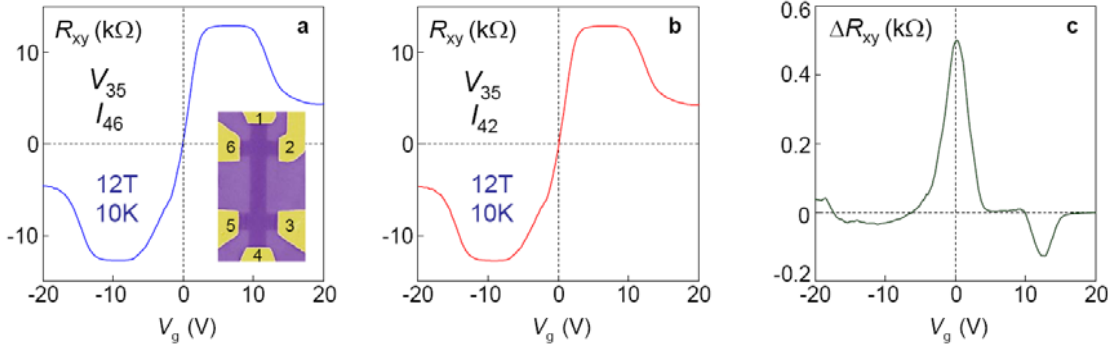

Fig. S1. Nonlocality in local transport. **a,b** – two sets of Hall measurements under exactly the same conditions but with swapping one of the current contacts. **c** – The difference between the two measurements $\Delta R_{xy}$ follows the behavior of nonlocal resistance $R_{35,26}$. The presented data are for GSiO with $\mu \approx 7,000$ cm$^2$/Vs. For high-$\mu$ GBN, the difference typically reaches several k$\Omega$.

#### #2. Graphene-on-BN devices

Graphene devices with $\mu \sim 10,000$ cm$^2$/Vs are now widely available and, to emphasize that the observed nonlocality is a commonplace phenomenon, much of the data presented in the main text were taken for GSiO. Furthermore, devices with million-range mobility can be obtained by suspending graphene. However, it has proven extremely difficult to make suspended 4-terminal devices, which are required for nonlocal measurements (see, for example, refs. [S3,S4]). Most recently [S5], it was demonstrated that atomically flat hexagonal boron-nitride (hBN) can be used as a quality inert substrate, which allowed devices with $\mu \approx 60,000$ cm$^2$/Vs, that is, three times higher than usually achievable for GSiO.

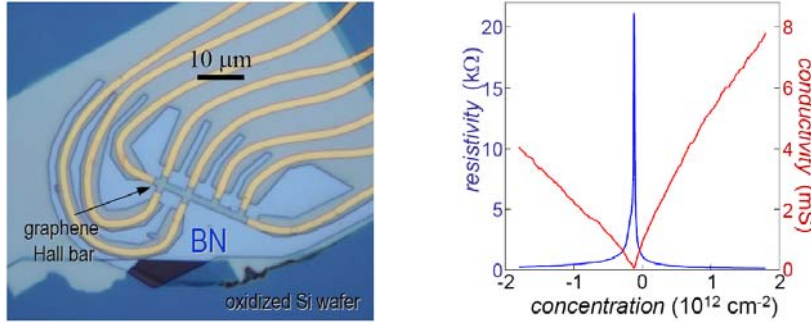

Fig. S2. Left – Hall bar made from graphene deposited on top of hBN [S5,S6]. hBN is $\approx 30$ nm thick and is residing on top of a Si wafer (90 nm of SiO$_2$). The image shows the device before the final step of removing a PMMA mask used for oxygen plasma etching. Right – Zero-*B* characteristics of one of GBN devices with $\mu \sim 50,000$ cm$^2$/Vs; $T = 60$ K.

In this work, we also used GBN devices, which allowed us to elucidate the scale of the observed nonlocality and better understand the physics underpinning this phenomenon. Our exfoliation and identification procedures for hBN are described in Ref. [S6]. Following the same extra steps in preparation procedures as described in Ref.

[S5], we have succeeded in making GBN devices with µ up to ≈150,000 cm$^2$/Vs. This refers to carrier concentrations $n$ between $10^{10}$ to $10^{11}$ cm$^{-2}$ (most of our GBN devices exhibited µ in the range from 50,000 to 100,000 cm$^2$/Vs). At higher $n$, µ gradually decreased which can be described by a short-range resistivity term $\rho_S$ of ~100Ω [S7], which varied for different devices and with $T$. Otherwise, the long-range mobility $\mu_L$ [S7] remained constant up to $n$ ~ a few $10^{12}$ cm$^{-2}$. Our GBN devices had little extrinsic doping ($10^{10}$ to $10^{11}$ cm$^{-2}$) and exhibited very high homogeneity such that, at low $T$, the Dirac point was smeared on a scale of only $n_0 \approx 10^{10}$ cm$^{-2}$.

#### #3. Dependence of nonlocal resistance on contact configuration

We have found that the nonlocality is strongly dependent on the exact contact configuration and usually changes for the opposite directions of $B$. Fig. S3 shows examples of $R_{NL}$ for several contact configurations. Generally, $R_{NL}$ becomes smaller as the distance between voltage and current probes $L$ increases and in the presence of extra leads between them (Fig. S3a). This data, however, is not sufficient to quantify the relaxation length $l$ involved in the nonlocal transport. Indeed, Fig. S3b shows nonlocal resistance measurements for the same sample and the same $L$ but with swapping current and voltage probes. One can see that $R_{NL}$ changes by more than a factor of 10 (red and black curves). Still, the Onsager relation holds as it should: $R_{35,26}(B) = R_{26,35}(-B)$ (see red and blue curves).

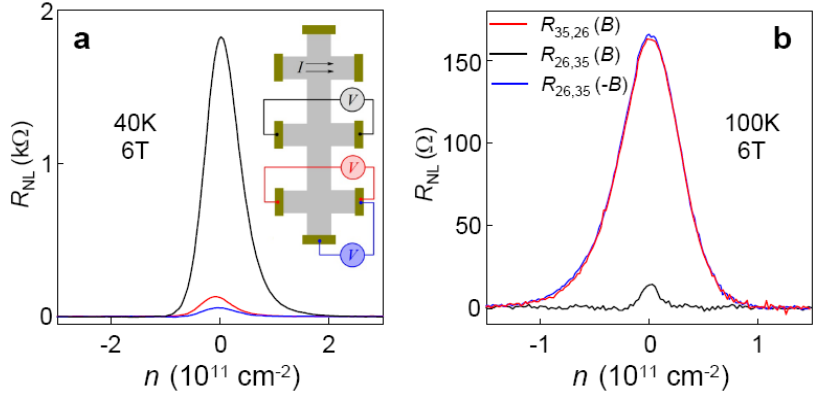

Fig. S3. Contact and sample dependence of the nonlocality. **a** – $R_{NL}$ measured for a GBN sample schematically shown in the inset. The curves are color coded: the current is applied through the top pair of contacts, whereas the voltage probe configurations are shown in the color corresponding to the black, red and blue curves. **b** – Nonlocal signal strongly varies from sample to sample. This was seen most clearly if we used the same contacts but swapped the current and voltage leads (numbers 2, 3, 5 and 6 refer to Fig. S1a). The signal can practically disappear for some geometries (black).

#### #4. Temperature and field dependence of the nonlocality in GSiO

Figures 2B-D of the main text plot the nonlocality in high-µ GBN where the amplitude of $R_{NL}$ reaches a value of ~1kΩ at room $T$. It is instructive to show that this behavior is generic and does not qualitatively change in the standard GSiO devices, neither in the quantum Hall effect (QHE) regime (Fig. 1 and 2A) nor in the quasiclassical regime. Fig. S4a is analogous to Fig. 2B of the main text. Both show essentially the same behavior but the $R_{NL}$ peak in GSiO is ~100 times smaller and twice wider than in GBN. The field dependence at high $T$ is monotonic for both GBN and GSiO (cf. Figs. 2D and S4b).

The qualitative difference between GBN and GSiO devices, which we have found, is their $T$ dependences (cf. Figs. 2C and S4c). Below 30K, GSiO exhibits a sharp rise in the nonlocal signal and, at intermediate $T$, $R_{NL}$ remains relatively constant. This behavior in GSiO is similar to the one observed in GBN and can again be attributed to the opening of a valley or many-body spin gap at low $T$. However, at higher $T$ (>100K in Fig. S4c), $R_{NL}$ in GSiO exhibits a rapid decay that is absent for GBN. We have found that the decay can be well fitted by a sum of two contributions, one is independent of $T$ and the other is thermally activated, $\propto \exp(-\Delta/T)$. The solid line in Fig. S4c is the best fit by a functional form $R_{NL} \propto 1/(\sigma_0 + \sigma_T \cdot \exp(-\Delta/T))$ where $\sigma_0$ an $\sigma_T$ could describe parallel channels of flavor relaxation. The fit in Fig. S4c yields an activation gap $\Delta \approx 1,000$K at 12T.

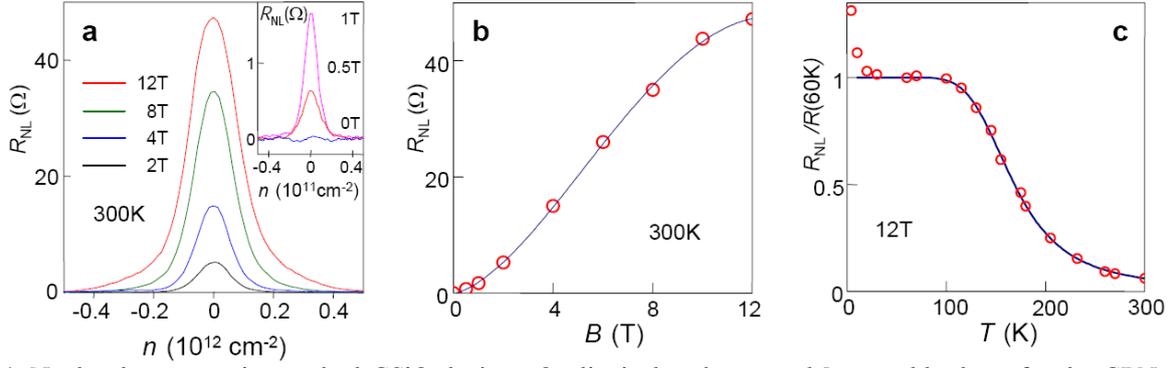

Fig. S4. Nonlocal transport in standard GSiO devices. Qualitatively, plots **a** and **b** resemble those for the GBN device shown in Figs. 2B and 2D of the main text, respectively. Note the scale of $R_{NL}$ which is now 100 times smaller. Nevertheless, the nonlocality is still easily detectable in $B>1$T at room $T$. **c** – $T$ dependence of $R_{NL}$ shows a much quicker decay of the nonlocality in GSiO as compared to GBN. This is attributed to an extra channel for spin flipping, which becomes dominant at elevated $T$ in GSiO. This GSiO device had $L \approx 5\mu$m and $w \approx 1\mu$m.

The $B$ dependence of $\Delta$ has been studied for 5 different devices. Figure S5 plots the inferred $\Delta$ in various $B$. One can see excellent reproducibility of the gap despite the absolute value of $R_{NL}$ varied strongly between the devices. Heuristically, we can describe the found dependence as $\Delta = v_F \cdot (2e\hbar B)^{1/2} - \Gamma$ (solid curve in Fig. S5) where the first term corresponds to the cyclotron gap between zero and first Landau levels ($v_F$ is the Fermi velocity in graphene; $e$ and $\hbar$ are the electron charge and the reduced Planck constant) and $\Gamma$ is the broadening of LLs. Typical $\Gamma$ found in our devices from the activation dependence between LLs are ~500K [S8], in agreement with the fit in Fig. S5, which yields $\Gamma \approx 400\pm100$K. This behavior can indicate the presence of an extra spin-flip process, which is responsible for the decay of $R_{NL}$ in GSiO and involves inter-LL scattering.

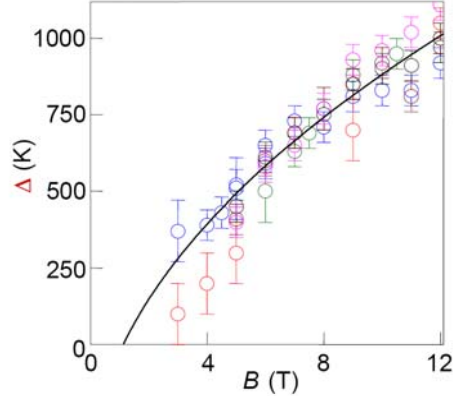

Fig. S5. Activation gap $\Delta$ inferred from $T$ dependence of nonlocal transport in GSiO at the NP in different magnetic fields.

#### #5. Absence of large spin/valley gaps at the NP

Magnetic field lifts the spin and/or valley degeneracy. Previous measurements [S3,S9] have shown that the flavor gaps $\delta$ are reasonably small and comparable in value with the Zeeman energy ($\approx15$K at 12T). However, transport phenomena in graphene can exhibit strong sample variations. Accordingly, we have also checked for the flavor gap in our samples. This was done by using the Corbino geometry. This geometry is necessary because spin splitting can lead to the dissipative quantum Hall effect with an insulating bulk and two counter-circulating edge states [S10,S11]. In the standard Hall bar geometry, this edge state transport electrically shots the bulk and does not allow to probe the spin gap as discussed in Ref. [S10,S11]. Figure S6 shows an example of our Corbino devices and a typical $T$ dependence of their 2-probe resistance in quantizing $B$. The $T$ dependence rules out any significant spin gap at the NP, which could otherwise explain the observed nonlocality by edge-state transport. The $T$ dependence at $\nu=2$ allows us to find the LL broadening $\Gamma \sim 500$K and to estimate the flavor gap at zero LL as δ ≤20K. The former agrees with the values reported in Ref. [S8] whereas the latter value is in agreement with the orthodox Zeeman splitting as well as measurements reported in Refs. [S3,S9,S12].

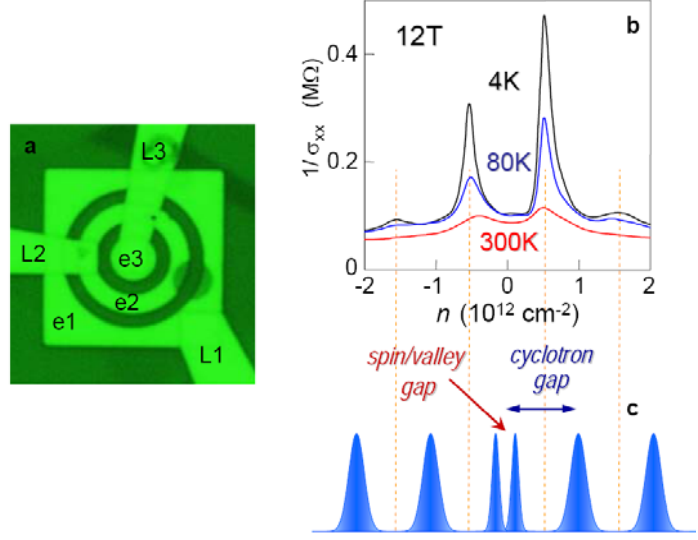

Fig. S6. Corbino measurements. **a** – Optical micrograph (using a green filter) of a Corbino device with three concentric electrodes (e1, e2 and e3) deposited on top of a graphene monolayer (GSiO). The electric leads are marked L1, L2 and L3. The scale is given by the 5μm diameter of the inner electrode e3. Leads L2 and L3 are electrically isolated from both graphene and the other electrodes by a layer of an electron-beam resist. **b** – Example of our Corbino measurements of longitudinal conductivity $\sigma_{xx}$ at different $T$. The magnetic field induces an insulating state at ν = 2 and 6 and leads to pronounced peaks in the 2-probe resistance between the Corbino electrodes. The gaps are illustrated schematically in **c**. Only a small rise in $\rho_{xx}$ (=1/$\sigma_{xx}$ at the NP) with decreasing $T$ could be seen near the NP, which rules out a large flavor gap.

#### #6. Nonlocal transport in the bulk or along edges?
A perfect zigzag edge in graphene presents a one-dimensional conductance channel with resistivity of $\sim h/e^2$. It is also predicted that a random edge can conduct electricity in a manner similar to zigzag [S13]. To asses the possibility that the observed nonlocality could be somehow mediated by an anomalously high conductivity of graphene edges, we have studied devices with widenings of the channel between current and voltage contacts. One of such devices is shown in Fig. S7. The micrograph shows a graphene mesa with several pairs of Hall contacts separated by approximately the same distance $L$ ~5μm. The conducting channel between the pairs could be either a straight ribbon or contain "bellies", that is, wider graphene regions in the middle. The bellies serve to increase the length of the edge between current and voltage probes in the nonlocal geometry. If the edges would be involved in the observed nonlocality, we should expect a strong suppression of $R_{NL}$ in the presence of the bellies. On the other hand, nonlocal currents mediated by the bulk are expected to be influenced much less by such bellies. We did not observe any significant difference in $R_{NL}$ for devices with and without bellies. This seems to rule out nonlocal transport mediated by graphene edges.

To further rule out a contribution of edge transport, we have performed a number of additional experiments. In one of them, we exposed a high-μ GBN device to $T$ above 300°C. This turned out to be detrimental for its electronic quality, reducing μ down to ~5,000 cm$^2$/Vs, presumably due to reaction of graphene with remnant air. The reduction in μ always resulted in strong suppression of the nonlocality (Fig. S8a). This behavior can be attributed to extra scatterers introduced in the graphene bulk, which reduces both μ and spin relaxation length. In another experiment, we fabricated side gates next to boundaries of a graphene Hall bar. These gates were made by etching narrow channels (~0.1 μm) within the same graphene crystal as shown in Fig. S8b. The central part of the crystal served as a multiterminal Hall bar device, whereas the periphery areas had independent contacts and could be used as side gates. Electrostatics modelling shows that the additional gates induced extra doping mostly near the edges with lesser influence in the bulk. Figure S8c shows $R_{NL}$ as a function of concentration $n$ (induced by the back gate) for two fixed side-gate voltages $V_{sg}$. The neutrality point could be shifted significantly by $V_{sg}$ (indicating a strip of extra doping near the edge) but we have found no notable difference in the strength of the nonlocality, which again is consistent with a bulk mechanism.

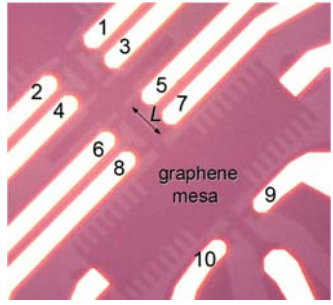

Fig. S7. Micrograph of a GSiO device made to probe the influence of graphene edges on nonlocal transport. The slightly darker areas are a graphene mesa. Bright areas are gold contacts. Configurations $R_{12,34}$ and $R_{56,78}$ provide the nonlocal measurements discussed in the main text. In the case of $R_{34,56}$ the current and voltage contacts are separated by the same distance $L$ as for $R_{12,34}$ but the channel contains a widening that increases the edge length. Very long edges are involved in the case of $R_{78,910}$.

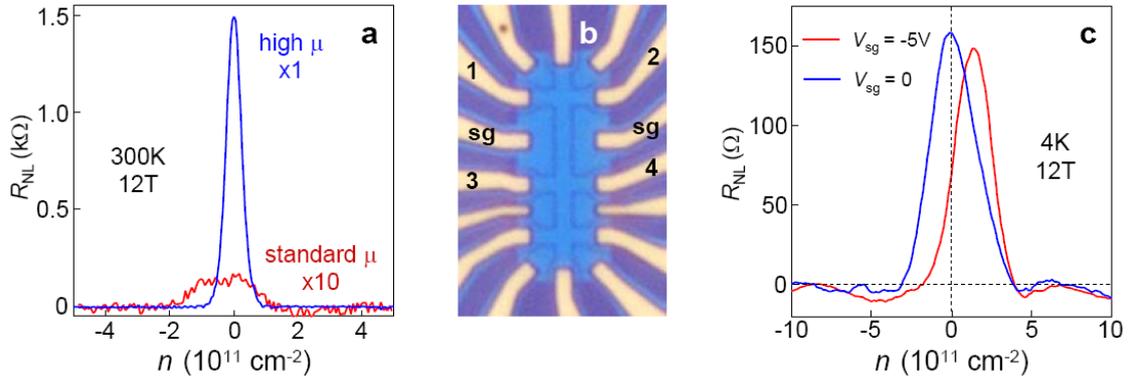

Fig. S8. Bulk vs edge. **a** – Decrease in graphene's electronic quality always results in weaker nonlocal signals. After µ was reduced by a factor of 30, we found a dramatic decrease in $R_{NL}$ (~100 times for the same distance between current and voltage contacts). **b** – Optical micrograph of a GSiO device with extra side gates. The light blue area is graphene under a layer of the resist used as an etch mask (it was removed later). **c** – $R_{NL} = R_{12,34}$ where the current and voltage probes are marked on the micrograph in (b). Side-gate voltage was applied to contacts marked as 'sg'. Except for the shift of the NP, the peak in $R_{NL}$ showed weak dependence on $V_{sg}$.

#### #7. Ohmic contribution to the nonlocal signal
In the main text, we have used the fact the ohmic contribution to the measured nonlocal signal becomes exponentially small when voltage contacts are positioned far away from the region of the classical current flow. In the strip geometry, this is describes by

$$R_{NL} \approx \frac{4}{\pi} \rho_{xx} \exp(-\pi L / w), \qquad L \gg w, \qquad (S1)$$

where $L$ is the distance between current and voltage probes, and $w$ is the strip width. The exponential dependence in this expression follows from the van der Pauw formula [S14],

$$\exp(-\pi R_{56,32} / \rho_{xx}) + \exp(-\pi R_{35,26} / \rho_{xx}) = 1.$$

It is instructive, however, to derive formula (S1) directly. In the derivation, we assume that the strip is situated at $-w/2 < y < w/2$, the source and drain contacts are positioned at $x = 0$, and the conductivity tensor is given by ($\sigma_{xx}$, $\sigma_{xy}$). The electric potential satisfies the Laplace equation, $\Delta\phi = 0$, as follows from the continuity equation $div(\vec{j}) = 0$, supplemented by the relations

$$\vec{j} = \hat{\sigma}\vec{E}, \quad \vec{E} = -grad(\phi),$$

where $\vec{E}$ is the electric field. The boundary conditions are given by $j_y(y = \pm w/2) = I_0 \delta(x)$ where the delta-function term models source and drain. Expressing current density in terms of potential, we obtain

$$\sigma_{xy}\partial_x\phi - \sigma_{xx}\partial_y\phi|_{y=\pm w/2} = I_0 \delta(x).$$

Solving the Laplace equation with the above boundary conditions, we find that the voltage drop $V$ a distance $L$ away from the source and drain is given by

$$V(L) = \phi(L,-w/2) - \phi(L,w/2) = 2I_0\rho_{xx}\int\frac{dk}{2\pi}\frac{e^{ikL}\sinh\frac{kw}{2}}{k\cosh\frac{kw}{2}}.$$

Evaluating the integral, we obtain

$$V(L) = \frac{4}{\pi}I_0\rho_{xx}\sum_{n=0}^{\infty}\frac{e^{-(2n+1)\pi L/w}}{2n+1} = \frac{I_0\rho_{xx}}{\pi}\ln\left[\frac{\cosh(\pi L/w)+1}{\cosh(\pi L/w)-1}\right],$$

which in the limit $L \gg w$ gives formula (S1). For typical experimental parameters $L/w = 5$ and $\rho_{xx} = 10$ k$\Omega$, we find $R_{NL} \sim 10^{-3}\Omega$, that is three orders of magnitude below the smallest nonlocal signal reported in our work.

#### #8. Spin Hall effect and nonlocal resistance in different regimes

Our model of nonlocal response mediated by spin diffusion, used to produce Fig. 3C of the main text, relies on the general approach developed in Ref. [S15]. It is assumed that, via spin-Hall effect (SHE), spin currents are generated by charge current passing through the system. The spin currents drive long-range spin transport, producing electric voltage at a remote location via inverse SHE. The relation between SHE and nonlocal resistance $R_{NL}$ is given by Eq. (12) of Ref. [S15]:

$$R_{NL} = \rho_{xx}\frac{we^{-L/l_s}}{2l_s}\left(\frac{\beta_s}{\sigma_{xx}}\right)^2, \qquad (S6)$$

where $\beta_s$ is the SHE coefficient that relates transverse spin current and electric field, and $l_s$ is the spin relaxation length. The meaning of the scales $L$ (length) and $w$ (width) is the same as in our current problem. In the situation analyzed in Ref. [S15] the SHE was of spin-orbital origin, whereas here we are interested in the SHE induced by interplay of the Zeeman interaction and magnetotransport, as described in the main text. However, the specifics of the microscopic origin of SHE do not impact the validity of Eq. (S6).

Therefore, we employ (S6) for calculating $R_{NL}$, using different models for the SHE coefficient $\beta_s$ and the charge resistivity tensor in the two regimes of interest: (i) the QHE regime, and (ii) the quasiclassical regime. The modeling procedure is summarized below for each of the two regimes. In both regimes, we will treat the factor $\frac{we^{-L/l_s}}{2l_s}$ in Eq. (S6), which contains unknown spin relaxation length $l_s$, as a fitting parameter.

First, we consider **the QHE regime**, which corresponds to a system with well-developed Landau levels. The QHE regime is realized at low $T$ and high $B$. In this case, we model the SHE neglecting the interactions between spin-up and spin-down carriers, and using the resistivity tensors defined independently for each spin projection (see Ref. [S16]). Then the fraction $\beta_s/\sigma_{xx}$ that enters Eq. (S6) should be replaced by the dimensionless SHE coefficient $\theta_{SH}/2$ defined in Ref. [S16] via the difference of the Hall angles for the two spin projections as

$$\theta_{SH} = \frac{\rho_{xy}^{\uparrow}}{\rho_{xx}^{\uparrow}} - \frac{\rho_{xy}^{\downarrow}}{\rho_{xx}^{\downarrow}} \qquad (S7).$$

Near the NP, where the strongest nonlocal response is observed, the main contribution to $\theta_{SH}$ arises from the difference of the Hall resistivities for carriers with opposite spin. Taking the longitudinal resistivities to be equal for both spin components, we approximate

$$\theta_{SH} \approx \frac{\rho_{xy}^{\uparrow} - \rho_{xy}^{\downarrow}}{2\rho_{xx}},$$

where the factor 2 is introduced to convert the resistivity for one spin projection to the total resistivity $\rho_{xx}$. To model the density dependence, we rewrite this formula in terms of the density of states $D$ and the Zeeman splitting $\delta$,

$$\theta_{SH} \approx \frac{\rho_{xy}^{\uparrow} - \rho_{xy}^{\downarrow}}{2\rho_{xx}} = \frac{1}{2\rho_{xx}} \frac{\partial \rho_{xy}}{\partial n} D\delta. \qquad (S8)$$

The density dependence of transport coefficients in the QHE regime is modeled following the approach described in Ref. [S17] (Gaussian broadening of LLs, and the semi-circle relation for the components of the conductivity tensor). In addition, we assume that the DOS is constant (smeared) in the vicinity of the Dirac point and treat its value $D$ as a fitting parameter. To fit the data for nonlocal response shown in Fig. 1C of the main text, we chose the parameter values $\lambda = 1.6$ (in notations of Ref. [S17]) and $D\delta/n_B \approx 0.2$, where $n_B$ is the particle density in a single LL. The value for the factor $\frac{we^{-L/l_s}}{2l_s}$ in Eq.(S6) was taken to be equal unity. The resulting density dependence of $R_{\text{NL}}$, shown in Fig. 3C of the main text, reproduces the key features of the data.

As a sanity check for our modeling procedure, we estimate the width $\Gamma$ of broadened zeroth LL, which corresponds to our choice of fitting parameters. Taking $\delta = 10K$ for Zeeman interaction at $B=12T$, we obtain $D = 0.2 n_B /\delta \approx 5 D_0$, where $D_0 = 4n_B / E_{01}$ is the density of states of a "fully smeared" LL (which corresponds to all the states in the four-fold degenerate LL smeared uniformly over the energy interval given by the cyclotron energy $E_{01} = \sqrt{2}\hbar v_F / l_B$). Thus, the chosen best-fit value for $D$ corresponds to five-fold enhancement of the density of states near the DP. This gives the LL width of about 1/5 the cyclotron energy $E_{01}$. For $B=12$ T, we obtain $\Gamma \approx 200K$, a reasonable value for GSiO.

We also comment on the maximum value of nonlocal resistance in the QHE regime. Typical value for the resistivity at the Dirac point is $\rho_{xx} = h/2e^2$, and the maximum possible value of the SHE coefficient $\theta_{SH}^{\max} \sim 2$ (which corresponds to $\rho_{xy}^{\uparrow(\downarrow)} \sim \pm \frac{h}{e^2}$). Then, taking the value $\frac{we^{-L/l_s}}{2l_s} = 1$ gives an estimate $R_{\text{NL}} = h/4e^2$, which is the value quoted in the main text. This value is in agreement with our measurements on GBN samples, where strong nonlocal response was observed.

***Quasiclassical regime*** – While the nonlocal response is found to be strongest in the QHE regime, our experiments show that nonlocality persists up to room $T$ and down to small $B$. In this regime, where the thermal broadening of LLs exceeds the LL separation, the notion of discrete LLs becomes invalid and, instead, a quasiclassical approach should be used.

In the quasiclassical regime, we use four-component transport model which takes into account both the disorder scattering and electron-hole scattering due to Coulomb interactions. The ratio $\beta_s / \sigma_{xx}$ that enters formula (S6) is then replaced by the dimensionless SHE coefficient $\xi_{SH}/2$ which is introduced and analyzed in the Appendix C of Ref. [S16]. We model the density dependence of transport coefficients using the procedure described in Ref. [S16]. The best agreement with the data on GBN samples at $B=1T$ and $T=300K$ was found for the following parameter values: disorder scale $\gamma = 60K$, which corresponds to density inhomogeneity $n_0 \sim (\hbar v_F/\gamma)^2 \approx 4 \cdot 10^9$

cm$^{-2}$, and the drag coefficient appropriate for BN, $\eta = 1.15\hbar$. The value for $\frac{we^{-L/l_s}}{2l_s}$ was chosen to be 1.6 to match the measured value of nonlocal resistance $R_{NL}$ at the NP. The resulting modelled density dependence of $R_{NL}$, shown in the inset of Fig. 3C of the main text, is in qualitative agreement with $R_{NL}$ measured at room $T$ (see the inset of Fig. 2B of the main text).